\documentclass[amsmath,notitlepage,amssymb,aps,showkeys,floatfix,prd,a4paper,
  onecolumn,nofootinbib]{revtex4-2}
\usepackage{times,amsbsy,amsfonts,graphicx,float}
\usepackage{color,morefloats,rotating,srcltx,slashed}
\usepackage{multirow,bm,verbatim,tabularx,bbding,threeparttable}
\definecolor{dblue}{rgb}{0.00,0.00,0.75}
\usepackage[colorlinks,urlcolor=dblue,linkcolor=dblue,citecolor=dblue]{hyperref} 
\usepackage{setspace}
\allowdisplaybreaks[4]

\begin{document} 

%\title{Revisiting the $J/\psi$ decay into three vectors with generation of resonances} 

\title{$J/\psi $ decay to $ \phi,\omega, K^{*0}$ plus $f_0(1370)$, $f_0(1710)$, $K_0^*(1430)$, $f_2(1270)$, $f'_2(1525)$ and $K_2^*(1430)$: role of the $D$-wave for tensor production}

\author{Luciano M. Abreu$^{1,2}$}     \email{luciano.abreu@ufba.br}
\author{Lianrong Dai$^{3}$}     			\email{dailianrong@zjhu.edu.cn}
\author{Eulogio Oset$^{2,4}$}         \email{oset@ific.uv.es}

\affiliation{$^1$Instituto de F\'{\i}sica, Universidade Federal da Bahia, Campus Universit\'{a}rio de Ondina, 40170-115 Bahia, Brazil\\
                 $^2$Instituto de F\'isica Corpuscular, Centro Mixto Universidad de Valencia-CSIC, Institutos de Investigaci\'on de Paterna, Aptdo. 22085, 46071 Valencia, Spain \\    
                 $^3$School of Science, Huzhou University, Huzhou 313000, Zhejiang, China \\
                 $^4$Department of Physics, Guangxi Normal University, Guilin 541004, China
}

%XXXXXXXXXXXXXXXXXXXXXXXXXXXXXXXXXX%
\begin{abstract}

We reassess the decay of the $J/\psi $ into an $\omega, \phi, K^{*0}$ and one of the $f_0(1370)$, $f_0(1710)$, $f_2(1270)$, $f'_2(1525)$, $K_0^*(1430)$ and $K_2^*(1430)$ resonances. We benefit from previous works that considered this reaction as a $J/\psi $ decay into three vector mesons, with a scalar or tensor resonance being formed from the interaction of two of these vectors. The novelty here with respect to former studies is the investigation of the relation between the scalar meson and tensor productions for the first time. To this end, the spin structure of the four vectors present in the production vertex is analyzed, and the $D$-wave mechanism in the tensor production is included. Then, beyond the ratios studied previously involving scalar states and tensor states independently, new ratios relating the scalar and tensor meson productions are estimated. Our results suggest that the $D$-wave mechanism of tensor production assumes a relevant contribution. New experimental data reporting the angular distributions of these processes will be important for checking this conclusion.

\end{abstract}

\date{\today}

%\pacs{13.20.He, 13.25.Hw, 13.30.Eg}
\maketitle
%XXXXXXXXXXXXXXXXXXXXXXXXXXXXXXXXXX% @ Begin

%%%%%%%%%%%%%%%%%%%%%%%%%%%%%%%%%%
\section{Introduction}                  %%%========SEC1
%%%%%%%%%%%%%%%%%%%%%%%%%%%%%%%%%%

The generation of resonances from the interaction of elementary hadrons occupies today a special place in Hadron Physics, from the pioneering work on meson-baryon interaction \cite{kaiser} and on meson-meson interaction \cite{oller} till now, when the unprecedented large amount of new states found experimentally has spurred a wave of research showing that many of the states found can be explained from this molecular hadron-hadron perspective (see reviews on the subject \cite{Oller:2000ma,Chen:2016qju,Hosaka:2016pey,Chen:2016spr,Lebed:2016hpi,Esposito:2016noz,Oset:2016lyh,Guo:2017jvc,Ali:2017jda,Olsen:2017bmm,Karliner:2017qhf,Yuan:2018inv,Liu:2019zoy,dongguo,dongzou,dong}). 

In the work of \cite{oller} (see also subsequent works \cite{Kaiser:1998fi,Locher:1997gr,Nieves:1999bx}) the interaction of pseudoscalar mesons was shown to produce the  $f_0(500), f_0(980), a_0(980)$ and $K^*_0(700)$ resonances and many reactions are explained from this perspective \cite{Oller:2000ma}.  The idea was extended to the interaction of pseudoscalar-vector interaction in~\cite{Lutz:2003fm,Roca:2005nm}, where it was shown that the axial-vector low lying resonances were also dynamically generated in this way. The next step was to consider the interaction of vector mesons among themselves, for which one had to shed aside the chiral Lagrangians \cite{Gasser:1983yg} and use the local hidden gauge approach  \cite{Bando:1987br,Harada:2003jx,Meissner:1987ge,Nagahiro:2008cv}. With this framework the $\rho \rho$ interaction was studied in \cite{Molina:2008jw}, where the $f_2(1270),f_0(1370)$ were dynamically generated, and the work was extended to SU(3) in \cite{Geng:2008gx} in which many other resonances were generated, among them the $f_0(1710), f'_2(1525), K_0^{\ast}(1430)$ and $K_2^{\ast}(1430)$. 

  The large binding of the $f_2(1270)$ stems from a very strong interaction in the $S$-wave with spins coupled to total angular momentum $J=2$. Because of this large binding, the molecular nature of this resonance was questioned in \cite{gulmez,guogul}, but the problems posed were solved in \cite{raquelgeng,ragengptep} confirming the molecular nature of the resonance. One recent push for the idea of \cite{Molina:2008jw,Geng:2008gx} came from the discovery of a new $a_0$ resonance in BABAR \cite{BaBar:2021fkz} and BESIII \cite{BESIII:2020ctr,BESIII:2021anf}, around 1710-1813 MeV, which had been predicted in \cite{Geng:2008gx} and later corroborated in \cite{guogul}. Some implications and perspectives of this new resonance are discussed in \cite{Abreu:2023xvw,gengdai}, and consistency of the properties of the  $a_0$ resonance found in \cite{Geng:2008gx} with the experimental findings of BESIII and BABAR is established in \cite{daiconsi,Zhu:2022wzk,Zhu:2022guw}.

   The agreement of the molecular picture for these vector-vector states with abundant experimental information has been discussed in a series of papers. In \cite{Branz:2009cv} the $ \gamma \gamma $ decay rate of the $f_2(1710)$ is evaluated and found consistent with the PDG information \cite{Workman:2022ynf}. In \cite{MartinezTorres:2012du} it is suggested that the peak observed at the $ \phi \omega $ threshold in the $ \phi \omega $ mass distribution of the $ J/\psi \to \gamma \phi \omega $ decay \cite{BES:2006vdb} is a consequence of the $f_0(1710)$ resonance. Predictions for other decay modes, and rates for $f_0(1710)$ production in decays of other particles are done in \cite{Geng:2010kma,Yamagata-Sekihara:2010mia,Xie:2014gla,Dai:2015cwa,Molina:2019wjj,Ikeno:2019grj}.

    With this large amount of cases of conformity of the theory with experiment there is still a new test than can be done and which we face here. This is the decay of the $J/\psi $ into an $\omega, \phi, K^{*0}$ plus one of the $f_0(1710), f'_2(1525), K_0^{\ast}(1430)$ and $K_2^{\ast}(1430)$ resonances. This is not the first time that the problem is addressed. Indeed, in \cite{daifirst} the problem was addressed by writing the $\omega, \phi$ as a combination of an SU$(3)$ singlet and octet and then combining the two extra vectors forming the resonance as also a combination of the singlet and octet. The formalism was borrowed from the similar work studying the decay of $J/\psi$ into $\omega, \phi$ and two pseudoscalars that interacted to form the $f_0(980)$ and $a_0(980)$ resonances \cite{Roca:2004uc,Lahde:2006wr}, which was found equivalent to a different one used in  \cite{Meissner:2000bc}. In \cite{Dai:2015cwa} the formalism was extended to study similar decays from $\psi(nS)$ and $\Upsilon(nS)$ vector states. In \cite{Molina:2019wjj} the problem of the $J/\psi $ into an $\omega, \phi, K^{*0}$ plus one of the $f_0(1710), f'_2(1525), K_0^{\ast}(1430)$ and $K_2^{\ast}(1430)$ resonances was retaken changing the formalism to a different one, more elegant and intuitive, in which the nature of the $J/\psi $ as a singlet of SU$(3)$ was exploited. The $q \bar q$ matrix in SU$(3)$ is written in terms of vector mesons, $V$, and then the three structures of SU$(3)$ invariants $\langle VVV\rangle$, $\langle VV\rangle \langle V\rangle$ and $ \langle V\rangle\langle V\rangle\langle V\rangle$ are considered, with $\langle \cdot\cdot\cdot\rangle$  standing for the trace of these matrices.  The idea of writing the decay of $J/\psi$ for three particles in terms of SU$(3)$ invariants was borrowed from  Ref. \cite{Liang:2019vhf}, where
the $ J/\psi \to \eta (\eta^{\prime} h_1(1380) $ BESIII decay~\cite{BESIII:2018ede} was studied. In this latter reaction the basic ingredient is the $ J/\psi \to V V P $ ($P$ for pseudoscalar) transition, followed by $V P$ interaction that, according to \cite{Lutz:2003fm,Roca:2005nm} generates the axial vector mesons. Once again, considering the $ J/\psi $ as a SU$(3)$ singlet, two structures were used $\langle VVP \rangle$ and $\langle V V  \rangle \langle P \rangle$,  and the $\langle V V P \rangle$ structure was shown dominant. The dominance of the trace of three mesons has also been established in the $\chi_{c1}$ decay into $\eta \pi^+ \pi^-$~\cite{BESIII:2016tqo}, where the $ \langle P P P \rangle$ structure for primary production of three pseudoscalars in the $\chi_{c1}$ decay is followed by the interaction
of $P P$ pairs to produce the $a_0(980)$ and the $f_0(500)$ resonances~\cite{Liang:2016hmr}. Similarly, the $ \langle VVP \rangle$ structure has also been tested as the dominant one in the study of the $ \chi_{c J}$  decay into $\phi h_1(1380) $~\cite{BESIII:2015vfb}, where $ \phi $ is a spectator and the other $V P$ pair interacts to produce the $h_1(1380)$ \cite{Jiang:2019ijx}. In \cite{Molina:2019wjj} the $ \langle VVV \rangle $ structure in the primary production was assumed, followed by the interaction of the $V V$ pairs, together with the $ \langle VV \rangle \langle V \rangle $ structure which was shown to have a much smaller strength. 
In retrospective, we can justify these results on the large $ N_c $ counting. According to Ref.~\cite{Manohar:1998xv}, ``diagrams with two quark loops have two flavor traces, and are of order of unity, and in general, those with $r$ quark loops have $r$ traces, and are of order $N_c^{1-r}$." This means that the structure with $ \langle V \rangle \langle V \rangle \langle V \rangle $ is suppressed by an order of $N_c^{-2}$ with respect to the $ \langle VVV \rangle $  one. 

In the present work we retake this problem with the same formalism concerning the flavor combinations, but we address for the first time the relationship of the scalar meson production ($f_0,K_0^{\ast}$) with tensor production  ($f_2,K_2^{\ast}$). We find that, while the scalar meson production proceeds via $S$-wave, the tensor production requires a large $D$-wave. Unfortunately, there are no data that shows angular distributions from where we could test this conclusion, but the present findings should provide an incentive to look again into these decays, in view of better present and planned future statistical data samples.

%%%%%%%%%%%%%%%%%%%%%%%%%%%%%%%%%%
\section{Formalism}\label{sec:2}           %%%========SEC2 
%%%%%%%%%%%%%%%%%%%%%%%%%%%%%%%%%%

%%%%%%%%%%%%%%%%%%%%%% Fig-1
\begin{figure}[tbp]   %tbp H
  \centering
  \includegraphics[width=8cm]{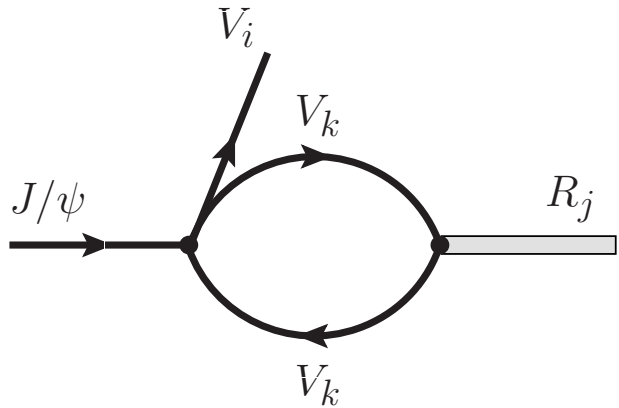}
  \caption{Mechanism for the production of the resonances $R_j$ from the $J/\psi$ decay into three vectors.
              }
  \label{fig1}
%\vspace{-0.2cm}
\end{figure}
%%%%%%%%%%%%%%%%%%%%%% Fig-1

The basic idea here is to take the $J/\psi$ as a singlet of SU$(3)$ and consider the process $J/\psi \to VVV$, and so the combination of $VVV$ must be a SU$(3)$ 
scalar; then a $VV$ pair should interact and produce scalar or tensor resonances. This reaction is depicted in Fig.~\ref{fig1}. As mentioned in the Introduction, we exploit this problem using the same formalism in Ref.~\cite{Molina:2019wjj}. Accordingly,  we write   the $q_i \bar q_j$ matrix ($i,j=u,d,s$) in terms of physical mesons
%%%%%%%%%%%%%%%
\begin{equation}
V = \left(
           \begin{array}{ccc}
             \frac{1}{\sqrt{2}}\rho^0 + \frac{1}{\sqrt{2}}\omega  & \rho^+ & K^{*+} \\
             \rho^- & -\frac{1}{\sqrt{2}}\rho^0 + \frac{1}{\sqrt{2}}\omega  & K^{*0} \\
            K^{*-} & \bar{K}^{*0} & \phi \\
           \end{array}
         \right),
   \label{eq4}
\end{equation}
%%%%%%%%%%%%%%%
and employ the SU$(3)$ invariant structure made from the product of $V$ matrices
\begin{equation}
 \langle VVV\rangle + \beta  \langle VV\rangle \langle V\rangle. 
 \label{vvvstructure}
\end{equation}
However, Ref.~\cite{Molina:2019wjj} ignored the spin structure coming from the four vectors present in the production vertex, and only calculated ratios between spin $J=0$ states and $ J=2 $ states independently, since the spin factors cancel in the ratios. Now we explore the spin structure, which allows us to relate the scalar  and tensor meson productions. To this end, we describe below the spin classification, and obtain the different contributions present in the invariant structures.

%%%%%%%%%%%%%%%%%%%%%%%%%%%%%%%%%%
\subsection{Spin classification}\label{sec:2.1}           %%%========SEC2.1 
%%%%%%%%%%%%%%%%%%%%%%%%%%%%%%%%%%

Since each vector in the reaction displayed in Fig.~\ref{fig1} carries a polarization vector $ \varepsilon $, we must construct a scalar amplitude by contracting the 
$ \varepsilon $'s. Working firstly in $S$-wave, a generic amplitude for the $J/\psi \to V_1 V_2 V_3 $  process should be written as 
\begin{eqnarray}
t & = & \tilde t_1 \,\varepsilon _{J/\psi} \cdot\varepsilon _1 \varepsilon  _2 \cdot\varepsilon _3 
+ \tilde t_2 \, \varepsilon_{J/\psi} \cdot\varepsilon _2  \varepsilon _1 \cdot\varepsilon _3  \nonumber \\
& & + \tilde t_3 \, \varepsilon_{J/\psi} \cdot\varepsilon _3   \varepsilon _1 \cdot\varepsilon _2 .
\label{ampA1}
\end{eqnarray}
We can decompose the polarization vectors in terms of the following spin projectors: scalar ($S=0$), antisymmetric tensor ($S=1$) and symmetric tensor ($S=2$)~\cite{Molina:2008jw}
, namely
\begin{eqnarray}
\mathcal{P} ^{(S=0) }  &  = & \frac{1}{3}\varepsilon _{J/\psi} \cdot\varepsilon _1 \varepsilon  _2 \cdot\varepsilon _3 , 
\nonumber \\
\mathcal{P} ^{(S=1) } &  = & \frac{1}{2}\left( \varepsilon_{J/\psi} \cdot\varepsilon _2  \varepsilon _1 \cdot\varepsilon _3   - \varepsilon_{J/\psi} \cdot\varepsilon _3   \varepsilon _1 \cdot\varepsilon _2 \right) \nonumber \\
\mathcal{P} ^{(S=2) }  &  = & \frac{1}{2}\left( \varepsilon_{J/\psi} \cdot\varepsilon _2  \varepsilon _1 \cdot\varepsilon _3   + \varepsilon_{J/\psi} \cdot\varepsilon _3   \varepsilon _1 \cdot\varepsilon _2 \right)-  \frac{1}{3}\varepsilon _{J/\psi} \cdot\varepsilon _1 \varepsilon  _2 \cdot\varepsilon _3. 
\label{ampA3}
\end{eqnarray}
Then, using this decomposition in Eq.~(\ref{ampA1}), the generic amplitude can be written as
\begin{equation}
t = (3\tilde t_1+\tilde t_2+\tilde t_3) \mathcal{P}^{(S=0)} + (\tilde t_2-\tilde t_3) \mathcal{P}^{(S=1)} + (\tilde t_2 + \tilde t_3) \mathcal{P}^{(S=2)}.
\label{ampA4}
\end{equation}
If these different contributions appear with equal weight, as it in our situation due to the symmetry of the vectors, then $\tilde t_1 = \tilde t_2 = \tilde t_3 \equiv \tilde t$, giving 
\begin{equation}
t = \left(  5 \,\mathcal{P}^{(S=0)}  + 2\, \mathcal{P}^{(S=2)} \right) \tilde t .
\label{ampA5}
\end{equation} 
Therefore, the spin factor in the amplitude must be 5 or 2 for a scalar or tensor resonance produced, respectively. 

%%%%%%%%%%%%%%%%%%%%%%%%%%%%%%%%%%
\subsection{Isospin states }\label{sec:2.2}  %%%========SEC2.2 
%%%%%%%%%%%%%%%%%%%%%%%%%%%%%%%%%%

The $VV$ states written in terms of the isospin states with $I=0$ are given by
\begin{eqnarray}
&&|K^*\bar K^*, I=0\rangle = \frac{-1}{\sqrt2}(K^{*+}K^{*-}+K^{*0}\bar K^{*0}),                          \nonumber\\
&&|\rho\rho, I=0, I_3=0\rangle = \frac{-1}{\sqrt6}(\rho^{+}\rho^{-}+\rho^{-}\rho^{+}+\rho^{0}\rho^{0}),                  \nonumber\\
&&|\omega\omega\rangle = \frac{1}{\sqrt2}|\omega\omega\rangle,                  \nonumber\\
&&|\phi\phi\rangle = \frac{1}{\sqrt2}|\phi\phi\rangle,                  \nonumber\\
&&|\rho \bar K^* , I=1/2, I_3=1/2 \rangle = \sqrt{\frac{2}{3}} |\rho^+ K^{*-}  \rangle - \sqrt{\frac{1}{3}} |\rho^0 \bar K^{*0}  \rangle ,    
\label{eqwfun} 
\end{eqnarray}
where the isospin multiplets are defined as $(K^{*+}, K^{*0})$, $(\bar K^{*0}, -K^{*-})$, $(-\rho^+,\rho^0,\rho^-)$; and the 
$\frac{1}{\sqrt2}$ extra factor is implemented because of the unitary normalization adopted for identical particles in the counting of states in the intermediate loops.

%%%%%%%%%%%%%%%%%%%%%%%%%%%%%%%%%%
\subsection{$ \langle VVV \rangle $ structure }\label{sec:2.3}  %%%========SEC2.3 
%%%%%%%%%%%%%%%%%%%%%%%%%%%%%%%%%%

Let us now evaluate the terms coming from the trace of the product $ V_1 V_2 V_3 $. 
When this calculation is performed keeping the order of the fields, and making explicit the polarization vectors of the vector fields, it can be seen that all the three possible orderings appear with equal weight due to the cyclical property of the trace.  
As a consequence, benefiting from the discussion done in section~\ref{sec:2.1}, and using the decomposition of the polarization vectors in spin projectors as well as the isospin states, we obtain the following contributions carrying the $ \omega, \phi $ and $ K^{*0} $ vectors: 

\begin{itemize}

\item $ \omega$ meson
\begin{equation}
\frac{3 \omega }{\sqrt{2}} \left(  5 \,\mathcal{P}^{(S=0)}  + 2\, \mathcal{P}^{(S=2)} \right) 
\left[ - \frac{3}{\sqrt{6}} \left( \rho \rho \right)^{(I=0)} - \frac{2}{\sqrt{2}} \left( K^* \bar K^* \right)^{(I=0)} + \frac{2}{\sqrt{2}} \left( \omega \omega \right)^{(I=0)}  \right],
\label{v3omega}
\end{equation} 
in which a $3!$ symmetry factor is taken into account for the $ \omega \omega \omega $ contribution;

\item $ \phi $ meson

\begin{equation}
\phi \left(  5 \,\mathcal{P}^{(S=0)}  + 2\, \mathcal{P}^{(S=2)} \right) 
\left[  \frac{3!}{\sqrt{2}} \left( \phi \phi \right)^{(I=0)} - \frac{3!}{\sqrt{2}} \left( K^* \bar K^* \right)^{(I=0)} \right], 
\label{v3phi}
\end{equation} 
where, again, a $3!$ symmetry factor is considered for the $ \phi \phi \phi $ term.

\item $ K^{*0} $ meson

\begin{equation}
3  K^{*0} \left(  5 \,\mathcal{P}^{(S=0)}  + 2\, \mathcal{P}^{(S=2)} \right) 
\left[ \sqrt{ \frac{3}{2}} \left( \rho \bar K^* \right)^{(I=1/2)} + \sqrt{ \frac{1}{2}} \left( \omega \bar K^* \right)^{(I=1/2)}  \right].
\label{v3kstar}
\end{equation} 

\end{itemize}

%%%%%%%%%%%%%%%%%%%%%%%%%%%%%%%%%%
\subsection{$ \langle VV \rangle \langle V \rangle $ structure }\label{sec:2.4}  %%%========SEC2.4
%%%%%%%%%%%%%%%%%%%%%%%%%%%%%%%%%%

We can estimate the weights for the $ \langle VV \rangle \langle V \rangle $ structure proceeding analogously as in the preceding subsection.Thus, the contributions carrying the $ \omega, \phi $ and $ K^{*0} $ vectors are: 

\begin{itemize}

\item $ \omega$ meson
\begin{equation}
\sqrt{2} \omega  \left(  5 \,\mathcal{P}^{(S=0)}  + 2\, \mathcal{P}^{(S=2)} \right) 
\left[ - \frac{3}{\sqrt{6}} \left( \rho \rho \right)^{(I=0)} - \frac{4}{\sqrt{2}} \left( K^* \bar K^* \right)^{(I=0)} + \frac{6}{\sqrt{2}} \left( \omega \omega \right)^{(I=0)} + \frac{2}{\sqrt{2}} \left( \phi \phi \right)^{(I=0)}  \right];
\label{v3omega2}
\end{equation}

\item $ \phi $ meson

\begin{equation}
 \phi  \left(  5 \,\mathcal{P}^{(S=0)}  + 2\, \mathcal{P}^{(S=2)} \right) 
\left[ - \frac{3}{\sqrt{6}} \left( \rho \rho \right)^{(I=0)} - \frac{4}{\sqrt{2}} \left( K^* \bar K^* \right)^{(I=0)} + \frac{6}{\sqrt{2}} \left( \omega \omega \right)^{(I=0)} + \frac{2}{\sqrt{2}} \left( \phi \phi \right)^{(I=0)}  \right];
\label{v3phi2}
\end{equation}

\item $ K^{*0} $ meson

\begin{equation}
2  K^{*0} \left(  5 \,\mathcal{P}^{(S=0)}  + 2\, \mathcal{P}^{(S=2)} \right) 
\left[ \sqrt{2} \left( \omega \bar K^* \right)^{(I=1/2)} + \left( \phi \bar K^* \right)^{(I=1/2)}  \right].
\label{v3kstar2}
\end{equation} 

\end{itemize}

Note that we can also combine $\omega$ or $\phi$ with $K^{*}$ instead of $\bar K^{*}$
to produce the complex conjugate resonance. Hence we should compare to the branching ratios of $(K^{*0} \bar{K}_0^{*0}(1430) + cc )$, etc., which are what are given in the PDG.  

%%%%%%%%%%%%%%%%%%%%%%%%%%%%%%%%%%
\subsection{Transition matrix: $S$-wave production}\label{sec:2.5}  %%%========SEC2.5 
%%%%%%%%%%%%%%%%%%%%%%%%%%%%%%%%%%

The analytical expression of the transition matrix associated to the mechanism depicted in Fig.~\ref{fig1} can then be written as 
\begin{eqnarray}
 t(J/\psi \to V_i R_j) = W^{(J)} \sum_k \left( h_{V_i , k} + \beta h_{V_i , k} ^{\prime} \right) \, g_{R_j,k} \, G_k(m_{R_j}) \,   
            \label{tranmatrix}
\end{eqnarray}
where $V_i $ is the vector in the final state $( V_i =  \omega, \phi , K^{*0} )$; $R_j$ accounts for the different resonances produced; $ W^{(J)} $ is the spin factor for the resonance $R_j$, which assumes the value $W^{(J=0,2)} =5,2 $, respectively; the sum over $k$ runs for the possible intermediate $VV$ channels associated to the corresponding final state, i.e. $k= \rho \rho ,  K^* \bar K^* , \omega \omega , \phi \phi $ for  $ V_i =  \omega, \phi $, and $k=   \rho \bar K^*, \omega \bar K^*,  \phi \bar K^* $ for  $ V_i =  K^{*0} $; $ h_{V_i , k} $ and 
$ h_{V_i , k} ^{\prime}  $ are the weights calculated in subsections \ref{sec:2.3} and \ref{sec:2.4}, and summarized in Table~\ref{weights}; $\beta$ is a free parameter determined by fitting the experimental data. Finally, $ G_k(m_{R_j}) $ is the loop function of the two intermediate vector mesons at resonance mass $ m_{R_j} $, and $ g_{R_j,k} $ is the coupling of the resonance $R_j$ to the channel $k$. The values of $ G_k(m_{R_j}) $ and $ g_{R_j,k} $ are calculated in Ref.~\cite{Geng:2008gx}, and for the resonances $f_0(1370), f_0(1710), f'_2(1525)$ and $\bar K_2^{*}(1430)$ are summarized in Tables I and II of~\cite{Molina:2019wjj} , together with their respective uncertainties. With respect to the resonance $\bar K_0^{*}(1430)$, its mass and couplings are from Table III of~\cite{Geng:2008gx}, and the $ G_k(m_{ K_0^{*}}) $ has been calculated using dimensional regularization with the subtraction constant $a=-1.85$; this information is available in Table~\ref{tableKstar0}.

%%%%%%%%%%%%-Table1
\begin{table}[!]
%\footnotesize
\centering
\caption{Summary of the weights  $ h_{V_i , k} $ and $ h_{V_i , k} ^{\prime}  $ calculated in subsections \ref{sec:2.3} and \ref{sec:2.4}. Here $k$ runs over the intermediate states $k= \rho \rho ,  K^* \bar K^* , \omega \omega , \phi \phi $ for  $ V_i =  \omega, \phi $, and over $k=   \rho \bar K^*, \omega \bar K^*,  \phi \bar K^* $ for  $ V_i =  K^{*0} $. }
\label{weights}
\setlength{\tabcolsep}{8pt}
\setstretch{1.2}
\begin{tabular}{lcccc}
\hline 

\hline 
       &  $ \rho \rho$  & $K^* \bar K^*$  & $ \omega \omega$  & $ \phi \phi $ \\
\hline
 $h_{\omega,k}$ & $-\frac{3\sqrt{3}}{2}$ & $-3$ & $3$ & $0$ \\
 $h_{\phi,k}$ & $0$ & $-3\sqrt{2}$ & $0$ &  $3\sqrt{2}$ \\
 $h_{\omega,k} ^{\prime}$ & $-\sqrt{3}$ & $-4$ & $6$ & $2$ \\
 $h_{\phi,k} ^{\prime}$ & $-\sqrt{\frac{3}{2}}$ & $-2\sqrt{2}$ & $\sqrt{2}$ &  $3\sqrt{2}$ \\
\hline 
       &  $\rho \bar{K}^*$  & $\omega \bar{K}^* $  & $\phi \bar{K}^* $ & \\
\hline
 $h_{K_0^{*},k}$ & $3\sqrt{\frac{3}{2}}$ & $\frac{3}{\sqrt{2}}$ & $3$ &  \\
 $h_{K_0^{*},k} ^{\prime}$ & $ 0 $ & $2\sqrt{2}$ & $2$ &  \\
%\hline
\hline
\end{tabular}
\end{table}
%%%%%%%%%%%%-Table1

%%%%%%%%%%%%-Table1
\begin{table}[!]
%\footnotesize
\centering
\caption{Values of the couplings  $g_{K_0^{*},k}$ and loop function $G_k $ in different channels at the resonance energy for the $ K_0^* $.  The couplings  $g_{K_0^{*},k}$ of the $K_0^*$ into the different channels and the pole position $ (m,\Gamma)=(1643,47) \rm MeV$ are from Table III of Ref.~\cite{Geng:2008gx}.
             }
\label{tableKstar0}
\setlength{\tabcolsep}{8pt}
\setstretch{1.2}
\begin{tabular}{lccc}
\hline 
       &  $\rho \bar{K}^*$  & $\omega \bar{K}^* $  & $\phi \bar{K}^* $ \\
\hline
 $g_{K_0^{*},k}$ & $8102- i \, 959 $ & $1370- i\,146 $  & $-1518 + i \, 209$ \\
 $G_k (\times 10^{-3})$ &  $-10.82 $ & $-10.42 $  &  $-4.46 $ \\
%\hline
\hline
\end{tabular}
\end{table}
%%%%%%%%%%%%-Table1

%%%%%%%%%%%%%%%%%%%%%%%%%%%%%%%%%%
\subsection{Inclusion of $D$-wave mechanism for tensor production}\label{sec:2.6}  %%%========SEC2.6 
%%%%%%%%%%%%%%%%%%%%%%%%%%%%%%%%%%

So far, we have only considered the $L=0$ production mechanism, reflected in the momentum-independent weights in Eq~(\ref{tranmatrix}). Since $J/\psi , V_i$ are coupled to $J=0$ from the $ \varepsilon _{J/\psi} \cdot\varepsilon _{V_i} $ factor, in the case of  $ R_j $ with $ S=0 $ the amplitude must proceed in $ S $-wave. In the case of  $ R_j $ with $ S=2 $ we can have $ S $-wave, as we have already done, as well as $ D $-wave for parity reasons. However, since all the resonances $ R_j $ are constructed in $ S $-wave, the angular momentum must come from the momentum $  \vec{q} $ of the $ V_i =  \omega, \phi , K^{*0} $.
In this sense, the structure of a $L=2$ mechanism must be of the type
\begin{eqnarray}
\varepsilon _{J/\psi} \cdot\varepsilon _{V_i} \varepsilon_{2} ^j \varepsilon _3 ^l 
\left( q^j q^l - \frac{1}{3} \vec{q}^{\,2} \delta^{j l } \right), 
\label{ampd1}
\end{eqnarray}
with $ \vec{\varepsilon}_{2}, \vec{\varepsilon} _3 $ being the polarization vectors of the other two vector mesons that produce the resonance. Keeping in mind that the linear combinations of the tensor structure present in this expression engender the components of the spherical harmonics function for $L=2$, $Y_2^m$, then the transition matrix for  $D$-wave production can be written as 
\begin{eqnarray}
 \tilde{t}^{\prime}(J/\psi \to V_i R_j) \propto  \frac{ \vec{q}^{\,2}}{M_V^2}\varepsilon _{J/\psi} \cdot\varepsilon _{V_i}  Y_2^{\ast m} \vert J=2, m\rangle ,
\label{ampd2}
\end{eqnarray}
where the factor $ 1/M_V^2 \,(M_V = 800 \, \rm MeV )$ has been introduced due to dimensional reasons. Therefore, we get
\begin{eqnarray}
 \overline{\sum} \sum \vert \tilde{t}^{\prime} (J/\psi \to V_i R_j) \vert ^2 & \propto & \left( \frac{\vec{q}^{\,2}}{M_V^2} \right)^2  \overline{\sum} \sum  \sum_{m,m'}\left( \frac{1}{4\pi} \right) \int d\Omega \, Y_2^{ m^{\prime}}  Y_2^{\ast m} \langle J=2, m^{\prime}  \vert J=2, m\rangle \nonumber \\
 & = & \left( \frac{\vec{q}^{\,2}}{M_V^2} \right)^2  \left( \frac{5}{4\pi} \right) . 
\label{ampd3}
\end{eqnarray}
where $ \overline{\sum} , \sum$ denote the average over the $ J/\psi  $ polarization and the sum over the $ V_i $ polarization, respectively. Hence, the  squared modulus of the weighted transition amplitude for $J=2$ resonances with $D$-wave  production mechanism must be replaced according to 
\begin{eqnarray}
 \overline{\sum} \sum  \vert t \vert ^2 \to \overline{\sum} \sum  \vert t \vert ^2  +  \overline{\sum} \sum \vert t^{\prime} \vert ^2  , 
\label{ampd4}
\end{eqnarray}
with $t$ being the contribution from $S$-wave production mechanism given by Eq.~(\ref{tranmatrix}), and  
\begin{eqnarray}
\overline{\sum} \sum \vert t^{\prime} \vert ^2 & \equiv &  \gamma^2 \left( \frac{\vec{q}^{\,2}}{M_V^2} \right)^2  \left( \frac{5}{4\pi} \right) \vert \sum_k \left( h_{V_i , k} + \beta h_{V_i , k} ^{\prime} \right) \, g_{R_j,k} \, G_k(m_{R_j}) \vert^2 . 
\label{ampd5}
\end{eqnarray}
In the latter equation  we have introduced the free parameter $\gamma$ to be fixed by the experimental data. 

To match the angular momentum, in principle one could also have $L^{\prime}=2$ for the intermediate $VV$ components of the loop of Fig.~\ref{fig1} giving rise to the resonance $R_j$. While this possibility is not excluded, making the state with $J=2$ similar to the deuteron with a large $S$-wave function and a small $D$-wave component, the theoretical framework used in~\cite{Molina:2008jw,Geng:2008gx} relies exclusively on the $S$-wave. In analogy with the deuteron case we also assume the $D$-wave part to be very small. We can give an extra argument in favor of the $D$-wave in the vector $V_i$ instead of $V_k$ in Fig.~\ref{fig1}. The  $V_k$ states in Fig.~\ref{fig1} are bound, their momenta correspond to those in the wave function and are small, of the order of $100-200 \, {\rm MeV} /c$. On the other hand, let us take a particular case, $J/\psi \rightarrow \omega f_2(1270)$. The momentum of the $\omega$ is $1148 \, {\rm MeV} /c$. It is clear that with this large momentum there is room for a $D$-wave contribution as we have assumed.

We would also like to mention that we have averaged $ \vert t \vert ^2 $ and $ \vert t^{\prime} \vert ^2 $ over the three polarizations of the $J/\psi$, while in $e^+e^- $ experiments the $J/\psi$ would appear transversely polarized in the $e^+e^- $ direction. We should choose transverse polarizations but also average over the $e^+e^- $ directions, where the polarizations will change. Thus, averaging over all the  $J/\psi$ polarizations is a fair procedure, in particular in the evaluation of the ratios that we do. 

%%%%%%%%%%%%%%%%%%%%%%%%%%%%%%%%%%
\subsection{Widths and ratios}\label{sec:2.7}  %%%========SEC2.7
%%%%%%%%%%%%%%%%%%%%%%%%%%%%%%%%%%

The amplitudes calculated above will be used as input in the standard expression of the width for each $J/\psi$ decay into $V_i R_j$, 
\begin{eqnarray}
  \Gamma (J/\psi \to V_i R_j) =\frac{1}{8 \pi}\frac{1}{m^2_{J/\psi}} q \left( \overline{\sum}\sum|t_i|^2 \right) ,
      \label{width}
\end{eqnarray}
where
\begin{align}
   q &= \frac{\lambda^{1/2}(m_{J/\psi}^2,m^2_{V_i},m_{R_j}^2)}{2m_{J/\psi}} ;\label{mom} 
\end{align}
with $ m_{V_i} $ standing for the mass of the vector meson $V_i$. It should be stressed that for the scalar resonances only the amplitude related to the $ S $-wave production mechanism (Eq.~(\ref{tranmatrix})) is employed; while for tensor resonances the one with both $ S $-wave and $D$-wave production mechanisms (Eqs. (\ref{ampd4}) and (\ref{ampd5})) is considered.

Hence, the ratios to be estimated are 
\begin{eqnarray}
 R_1 & \equiv & \frac{\Gamma(J/\psi \to \phi f_2(1270))}{\Gamma(J/\psi \to \phi f_2^{\prime}(1525))}, \quad  R_2  \equiv \frac{\Gamma(J/\psi \to \omega f_2(1270))}{\Gamma(J/\psi \to \omega f_2^{\prime}(1525))}, \nonumber \\
 R_3 & \equiv & \frac{\Gamma(J/\psi \to \omega f_2(1270))}{\Gamma(J/\psi \to \phi f_2(1270))}, \quad  R_4  \equiv \frac{\Gamma(J/\psi \to K^{*0} \bar{K}_2^{*0}(1430))}{\Gamma(J/\psi \to \omega f_2(1270))}, \nonumber \\
 R_5 & \equiv & \frac{\Gamma(J/\psi \to \omega f_0(1370))}{\Gamma(J/\psi \to \omega f_0(1710))}, \quad  R_6  \equiv \frac{\Gamma(J/\psi \to \phi f_0(1370))}{\Gamma(J/\psi \to \phi f_0(1710))}, \nonumber \\
 R_7 & \equiv & \frac{\Gamma(J/\psi \to \omega f_0(1710))}{\Gamma(J/\psi \to \phi f_0(1710))}, \quad  R_8  \equiv \frac{\Gamma(J/\psi \to \omega f_0(1710))}{\Gamma(J/\psi \to \omega f_2(1270))}, \nonumber \\
 R_9 & \equiv & \frac{\Gamma(J/\psi \to \phi f_0(1710))}{\Gamma(J/\psi \to \phi f_2(1270))}, \quad  R_{10}  \equiv \frac{\Gamma(J/\psi \to K^{*0} \bar{K}_0^{*0}(1430))}{\Gamma(J/\psi \to K^{*0} \bar{K}_2^{*}(1430))}, \nonumber \\
 R_{11} & \equiv & \frac{\Gamma(J/\psi \to K^{*0} \bar{K}_0^{*0}(1430))}{\Gamma(J/\psi \to \omega f_2(1270))}.
 \label{eq_ratios}
\end{eqnarray}

In the next section we will present and discuss the results emphasizing the main novelties with respect to Ref.~\cite{Molina:2019wjj} and other previous works: the inclusion of the $D$-wave in the tensor production, together with the evaluation of the new ratios $R_{8}-R_{11}$ relating the scalar and tensor meson productions.

%%%%%%%%%%%%%%%%%%%%%%%%%%%%%%%%%%
\section{Results and discussions}            \label{sec:3}       %%%========SEC3
%%%%%%%%%%%%%%%%%%%%%%%%%%%%%%%%%%

%%%%%%%%%%%%%%%%%%%%%%%%%%%%%%%%%%

We start by revisiting the ratios $R_1-R_7$ already estimated in Ref.~\cite{Molina:2019wjj}. In this previous study  their dependence with the relative strength $\beta $ between the $\langle VVV\rangle $ and $ \langle VV\rangle \langle V\rangle $ structures has been investigated. Good agreement with experimental data was found with $\beta $ about $0.3$,  which would be in agreement with the expectations of the large $ N_c $ counting of~\cite{Manohar:1998xv} discussed in the Introduction. Anticipating different structures for $ J=0, J=2 $ states, only ratios between scalar and between tensor resonances independently were studied in previous works.  Here we are able to estimate the influence of $D$-wave mechanism for tensor production, encoded in the magnitude of the parameter $\gamma$. 

\begin{figure}
\centering
\includegraphics[width=0.48\columnwidth]{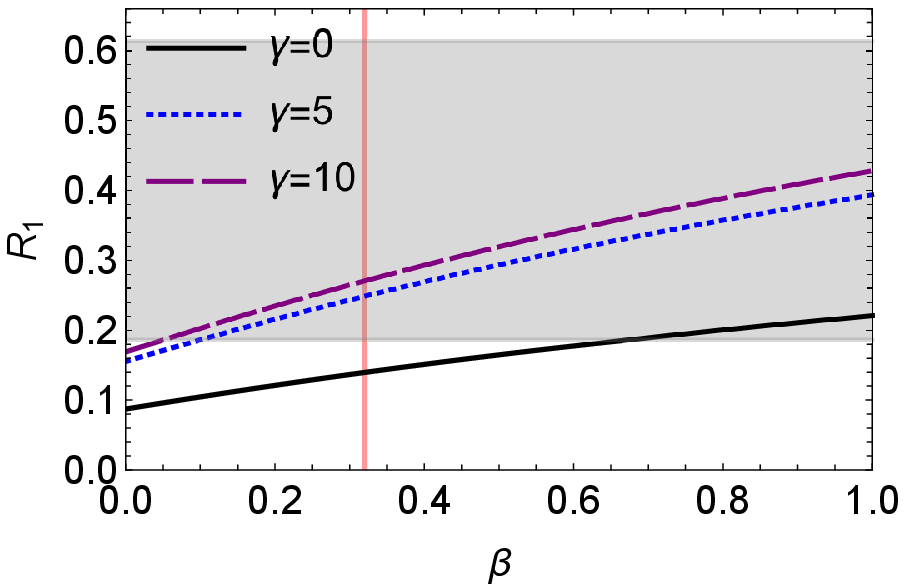}
\includegraphics[width=0.48\columnwidth]{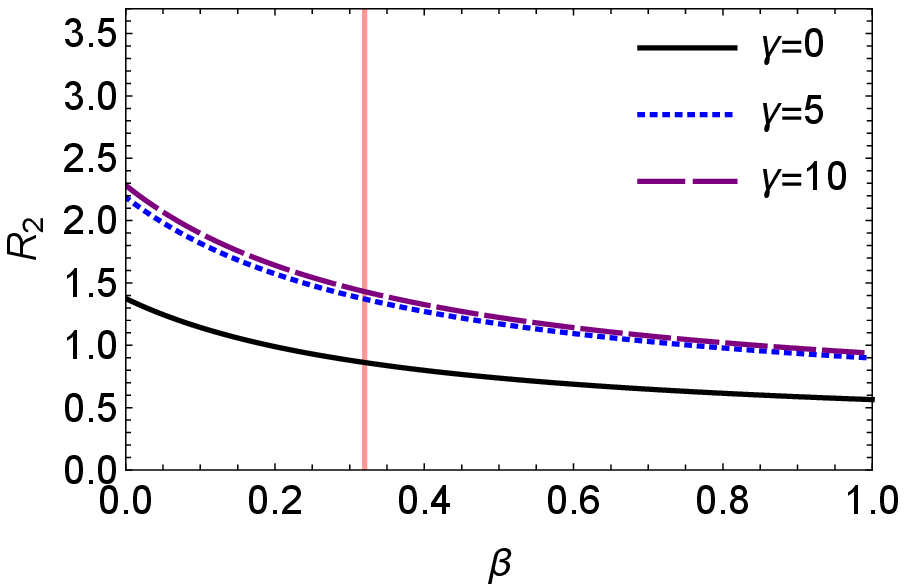} \\
\includegraphics[width=0.48\columnwidth]{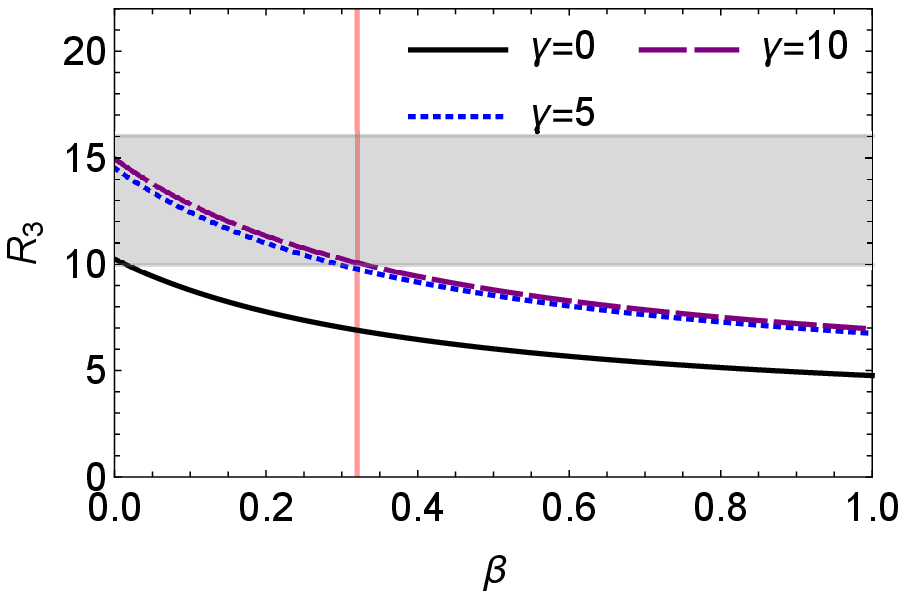}
\includegraphics[width=0.48\columnwidth]{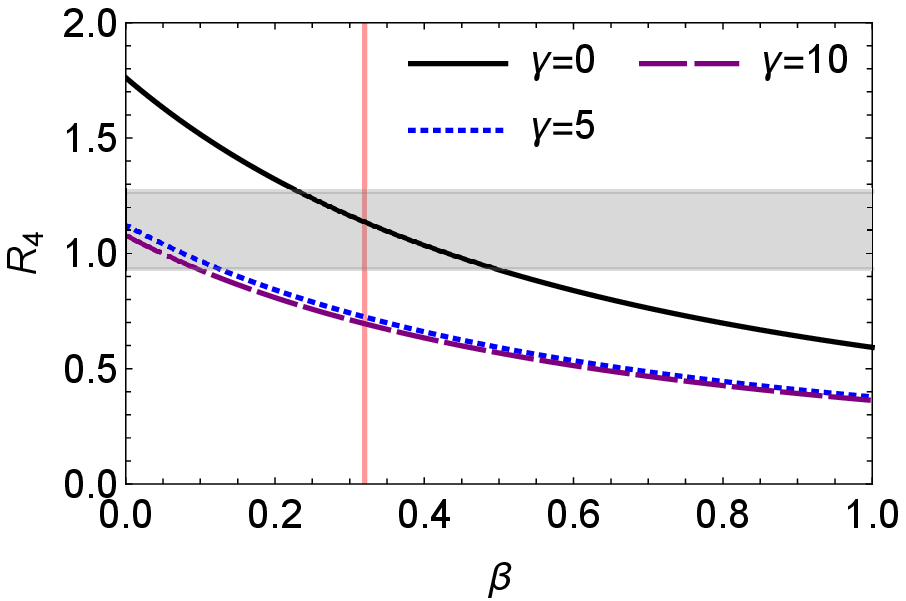} 
\caption{Dependence of the ratios $R_1-R_4$, defined according to Ref.~\cite{Molina:2019wjj},  with the parameter $ \beta $. The notation is the same as in Fig. 3 of Ref.~\cite{Molina:2019wjj}. But here only the central values of the ratios have been plotted. Gray bands represent the experimental result (when available), while the red vertical line indicates the value for $\beta =0.32$. As it can be seen, these results agree with those reported in~\cite{Molina:2019wjj}.
%In the case of $R_4$, the right panel takes into account the fact that $\Gamma (J/ \psi \to K^{*0} \bar{K}_2^{*}) ) = \dfrac{1}{2} \Gamma (J/ \psi \to K^{*0} \bar{K}_2^{*} + c c )$ in experimental result. 
}
\label{ratios1to4}
\end{figure}

\begin{figure}
\centering
\includegraphics[width=0.48\columnwidth]{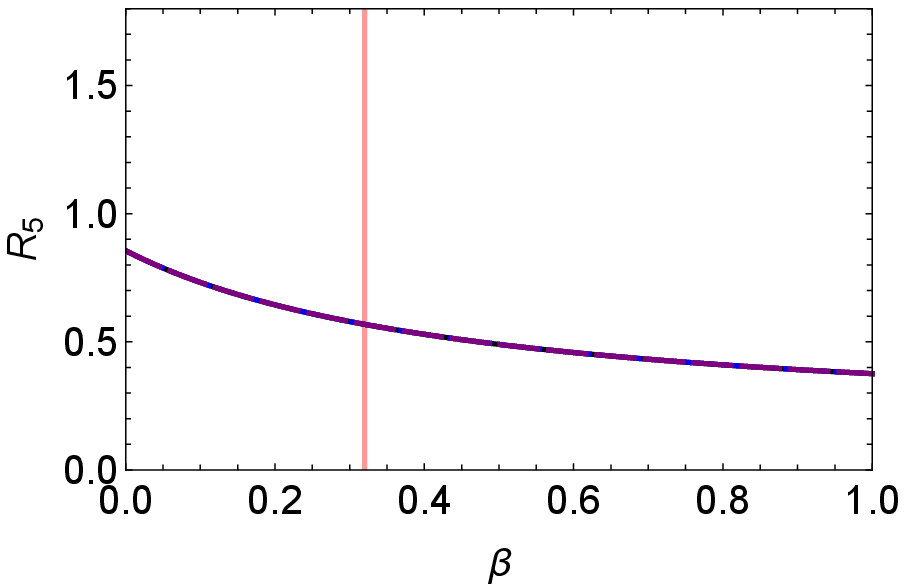}
\includegraphics[width=0.48\columnwidth]{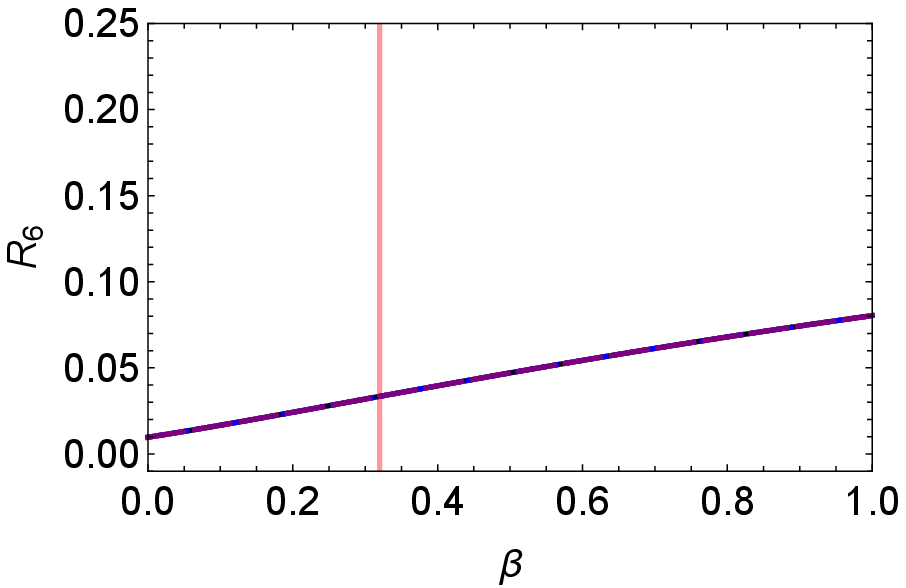} \\
\includegraphics[width=0.48\columnwidth]{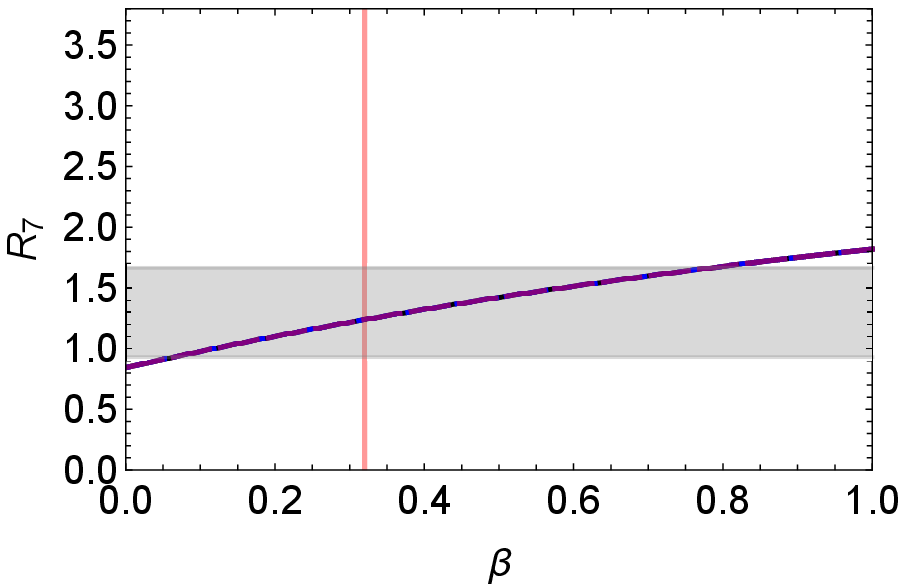}
\caption{Dependence of the ratios $ R_5-R_7 $ with the parameter $ \beta $. The notation is the same as in Fig. \ref{ratios1to4}. As expected, the curves for different values of $ \gamma $ coincide.}
\label{ratios5to7}
\end{figure}

So, in Figs.~\ref{ratios1to4} and \ref{ratios5to7} we show the dependence of the ratios $R_1-R_7$ defined in Eq.~(\ref{eq_ratios}) with $ \beta $, taking different values of the parameter $\gamma $.  The notation is the same as in Fig. 3 of Ref.~\cite{Molina:2019wjj}. In order to make the plot clearer, only the central values of the ratios are displayed. Gray bands represent the experimental data (available for $R_1, R_3, R_4$ and $R_7$), while the red vertical lines indicate $\beta =0.32$ used in~\cite{Molina:2019wjj}.  As expected, the results for  $\gamma = 0$ agree with those reported in~\cite{Molina:2019wjj}. 
%Notice that in the case of $R_4$, the left panel uses the same experimental range as in~\cite{Molina:2019wjj}, while the right panel takes into account the fact that $\Gamma (J/ \psi \to K^{*0} \bar{K}_2^{*}) ) = \dfrac{1}{2} \Gamma (J/ \psi \to K^{*0} \bar{K}_2^{*} + c c )$ in experimental result. 
Also, the curves for the ratios $ R_5-R_7 $ coincide when different values of $ \gamma $ are taken, since these ratios are related to reactions involving only scalar resonances, in which the $D$-wave mechanism does not contribute. On the other hand, ratios $R_1-R_4$ present sizeable dependence, with the $D$-wave playing a more important role as $\gamma$ increases. We note that higher values of $ \gamma $ engender even better accordance with experiments for $ R_1,R_3$. 

\begin{figure}
\centering
\includegraphics[width=0.48\columnwidth]{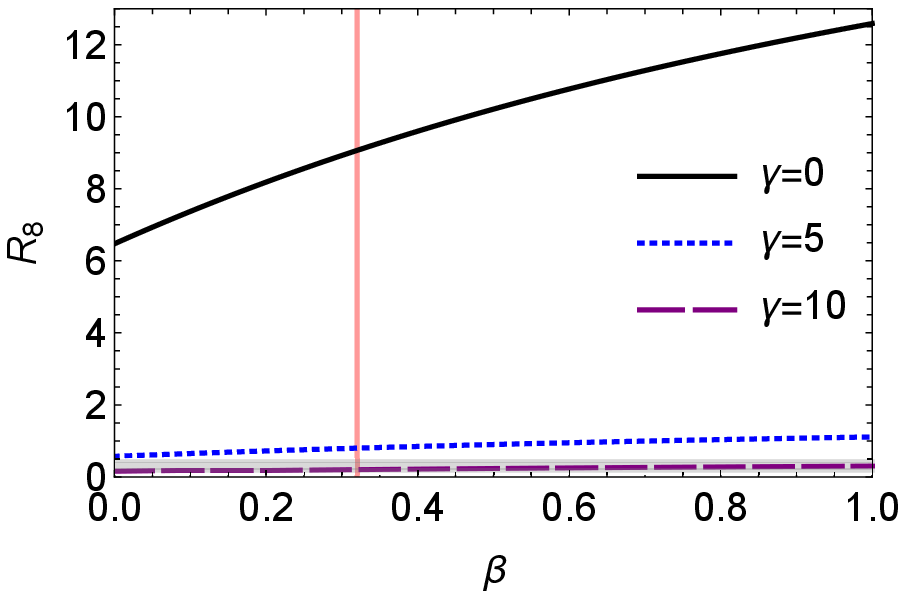}
\includegraphics[width=0.48\columnwidth]{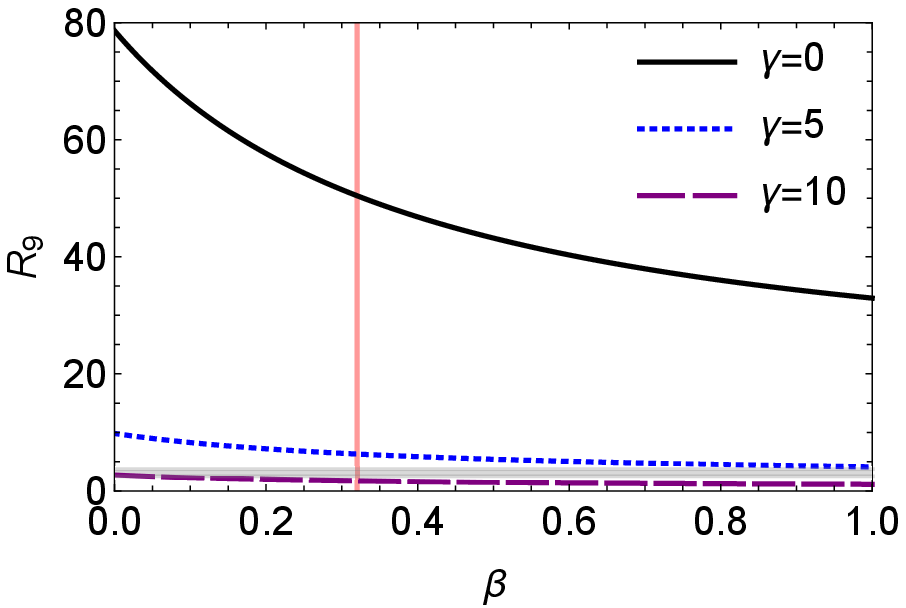}
\caption{Dependence of the ratios $R_8 $ and $ R_9$  with the parameter $ \beta $. The notation is the same as in Fig. \ref{ratios1to4}. 
}
\label{ratio89}
\end{figure}

Let us move on to the ratios $R_{8}-R_{11}$ relating the scalar and tensor meson productions, by considering firstly those with available experimental data. In Fig.~\ref{ratio89} we show the dependence of the ratios $R_8 $ and $ R_9$ with the parameter $ \beta $, taking different values of the parameter $\gamma $. We remark that the non-inclusion of $D$-wave mechanism in tensor production, i.e. $ \gamma = 0 $, yields severe disagreement of  $R_8 $ and $ R_9$ with respect to the experiments, by a factor of about $10-50$. Notwithstanding, when the $D$-wave mechanism is taken into account, we see that the augmentation of $ \gamma $ generates smaller values for these ratios closer to the experimental data, which suggests that this contribution assumes an important role in this scenario. 

\begin{figure}
\centering
\includegraphics[width=0.48\columnwidth]{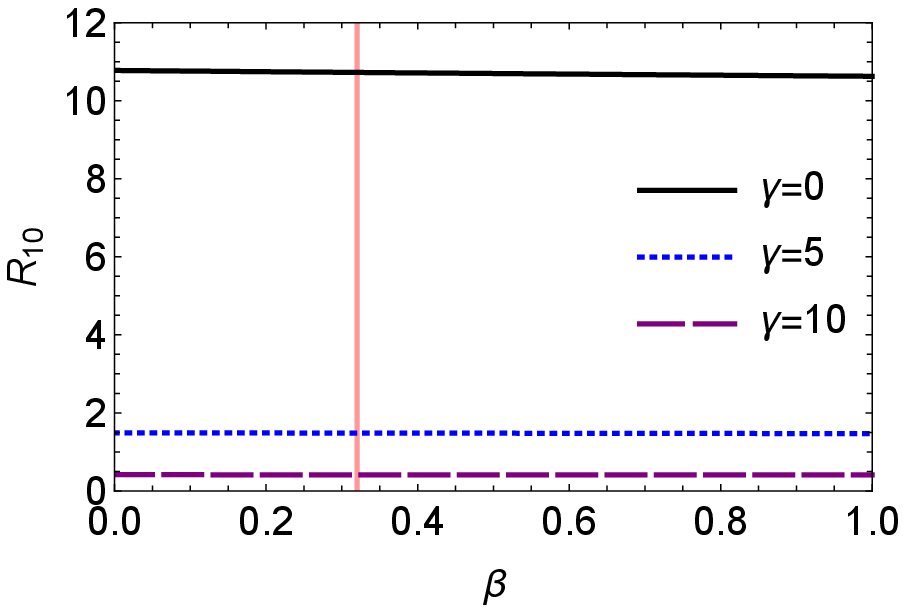}
\includegraphics[width=0.48\columnwidth]{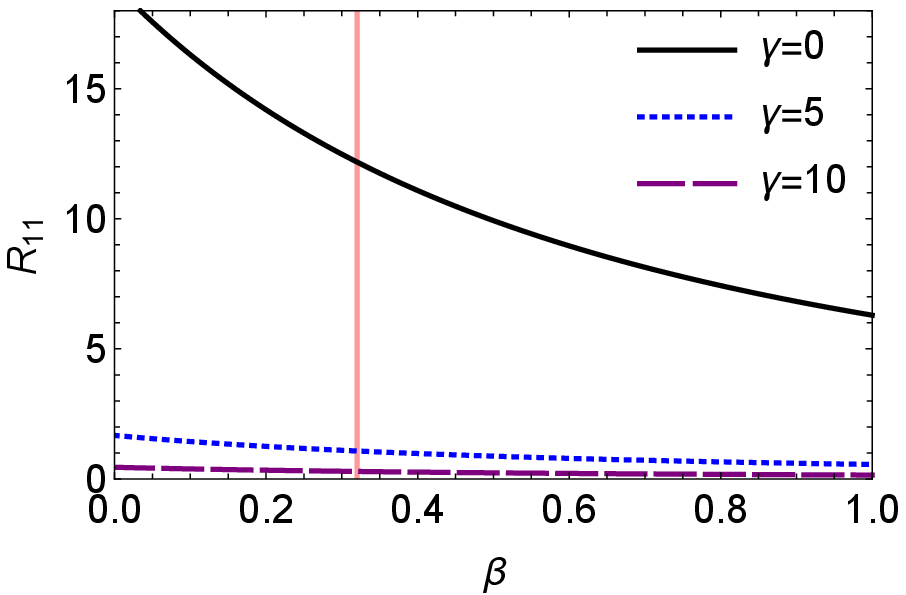}
\caption{Dependence of the ratios $ R_{10},  R_{11}$ with the parameter $ \beta $. The notation is the same as in Fig. \ref{ratios1to4}. 
%In Table \ref{table1} we show the values of the loop function in different channels at the resonance energy for the $ K_0^* $. 
}
\label{ratios1011}
\end{figure}

In Fig.~\ref{ratios1011} we plot the dependence of the ratios $R_{10} $ and $ R_{11}$ with the parameter $ \beta $, taking different values of the parameter $\gamma $. Looking at the case $\gamma =0$, the ratio $R_{10}$ has a mild dependence with $\beta$, whereas  $R_{11}$ suffers a significant change. The incorporation of $D$-wave mechanism in tensor production yields sizeable effects: similarly to ratios $R_8 $ and $ R_9$, $R_{10} $ and $ R_{11}$ decrease as $ \gamma $ growths, and the dependence of $ R_{11}$ with $\beta$ is softened. Unfortunately we cannot assess these findings for $R_{10} $ and $ R_{11}$ more concretely, since experimental data are not yet available. In this regard, these results are predictions to be checked by future experimental works.

In order to have a more detailed estimate of the $S$-wave and $D$-wave contributions for the reactions involving tensor resonances, in Table~\ref{fractionsd} we show the fractions of the $S$- and $D$-wave mechanisms for the branching ratios $( \Gamma_{i}^{S,D}/\Gamma_{i})$, taking different values of the parameter $\gamma $ for a value of $\beta=0.3$ (note that these rates are insensitive to $\beta$). In all cases, when $\gamma $ increases the contribution of the $D$-wave mechanism becomes dominant. 
%As stated before we take advantage of the previous work~\cite{Molina:2019wjj}, which pointed out reasonable conformity with data for the ratios $R_1 - R_7$ at $\beta \simeq 0.3$. 
Then, using this fact and taking into consideration the behaviour of the ratios with $\gamma $, in particular $R_8$ and $R_9$, the value of $\gamma \simeq 10$ appears as an appropriate choice. 
%Thus, in Fig.~\ref{fractionsd2} we summarize the fractions of the $S$- and $D$-wave mechanisms as functions of the parameter $ \beta $ with $\gamma = 10$. The $D$-wave mechanism contribution appears then large and predominant. 

%\begin{figure}
%\centering
%\includegraphics[width=0.48\columnwidth]{fractionsd1.eps}
%\includegraphics[width=0.48\columnwidth]{fractionsd3.eps} \\
%\includegraphics[width=0.48\columnwidth]{fractionsd4.eps}
%\includegraphics[width=0.48\columnwidth]{fractionsd5.eps} \\
%\includegraphics[width=0.48\columnwidth]{fractionsd6.eps}
%\caption{Dependence of the fractions $ \Gamma_{i}^{s,d}/\Gamma_{i}$ with the parameter $ \beta $, taking different values of the parameter $\gamma $.  $ \Gamma_{i}^{s}$ and $ \Gamma_{i}^{d}$ represent the contributions coming from the $s-$ and $d-$waves, respectively; and $i$ denotes the reactions involving tensor resonances, i.e. $\Gamma_{1} \equiv \Gamma (J/\psi \to \omega f_2(1270)), \Gamma_{3} \equiv \Gamma (J/\psi \to \omega f_2^{\prime}(1525)),\Gamma_{4} \equiv \Gamma (J/\psi \to K^{*} \bar{K}_2^{*}(1430) , \Gamma_{5} \equiv \Gamma (J/\psi \to \phi f_2(1270)), \Gamma_{6} \equiv \Gamma (J/\psi \to \phi f_2^{\prime}(1525))$. 
%}
%\label{fractionsd}
%\end{figure}

%\begin{figure}
%\centering
%\includegraphics[width=0.48\columnwidth]{fraction-s-d-wave.eps}
%\caption{Dependence of the fractions $ \Gamma_{i}^{s,d}/\Gamma_{i}$ with the parameter $ \beta $, taking  $\gamma = 10$.  The notation is the same as in Fig. \ref{fractionsd}. }
%\label{fractionsd2}
%\end{figure}

%%%%%%%%%%%%-Table1
\begin{table}[!]
%\footnotesize
\centering
\caption{Fractions of the $S$- and $D$-wave mechanisms for the branching ratios $( \Gamma_{i}^{S,D}/\Gamma_{i})$, taking different values of the parameter $\gamma $. }
\label{fractionsd}
\setlength{\tabcolsep}{8pt}
\setstretch{1.2}
\begin{tabular}{l|cc|cc|cc}
\hline 
       & $\gamma = 1 $ &  & $\gamma = 5 $ & & $\gamma = 10 $ &  \\
Reaction &  $ \frac{\Gamma^{(S)}}{\Gamma} $  &  $  \frac{\Gamma^{(D)}}{\Gamma} $  &  $ \frac{\Gamma^{(S)}}{\Gamma} $  &  $  \frac{\Gamma^{(D)}}{\Gamma} $   &  $ \frac{\Gamma^{(S)}}{\Gamma} $  &  $  \frac{\Gamma^{(D)}}{\Gamma} $   \\
\hline
$J/\psi \to \omega f_2(1270)$	&	0.71	&	0.29 	& 0.09	&	0.91  & 0.02	&	0.98 	 \\
$J/\psi \to \omega f_2^{\prime}(1525)$&	0.81	&	0.19 	& 0.14	&	0.86  & 0.04	&	0.96 	 \\
$J/\psi \to K^{*} \bar{K}_2^{*}(1430)$	&	0.80	&	0.20 	& 0.14	&	0.86  & 0.04	&	0.96 	 \\
$J/\psi \to \phi f_2(1270)$	&	0.78	&	0.22 	& 0.12	&	0.88  & 0.03	&	0.97 	 \\
$J/\psi \to \phi f_2^{\prime}(1525)$	&	0.88	&	0.12 	& 0.22	&	0.78  & 0.07	&	0.93 	 \\

%\hline
\hline
\end{tabular}
\end{table}
%%%%%%%%%%%%-Table1

\begin{figure}
\centering
\includegraphics[width=0.48\columnwidth]{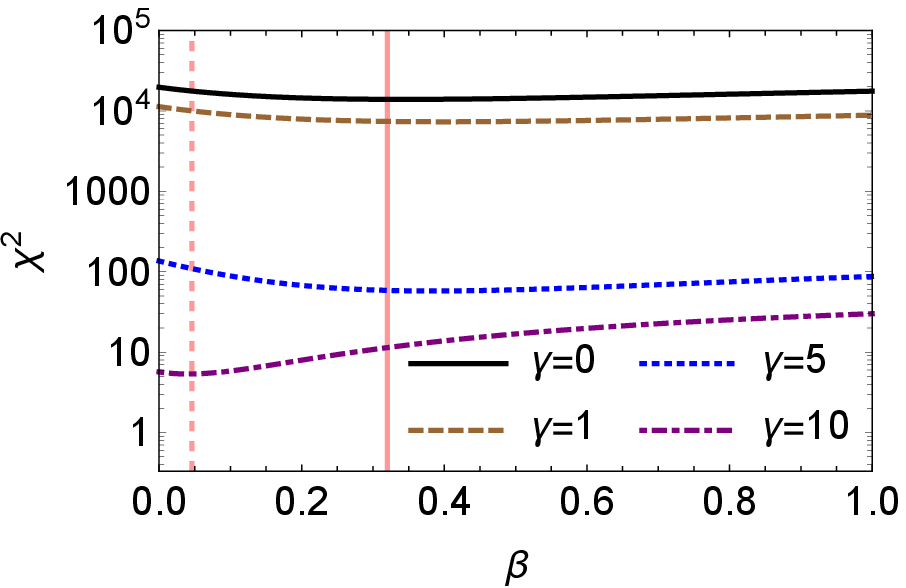}
\caption{Dependence of the function $\chi^2 $ with the parameter $ \beta $, taking different values of the parameter $\gamma $. 
%The dark curve represents the approach used in Ref.~\cite{Molina:2019wjj}.
% Inset: a zoom in the region of the curve with $\gamma =10$; the red vertical dashed line indicates the value for $\beta =0.21$. 
}
\label{chisquared}
\end{figure}

Let us also analyze how the $\chi^2 $ function changes with the free parameters of the theory. So, in Fig.~\ref{chisquared}, $\chi^2 $ is plotted as a function of $ \beta $, taking different values of the parameter $\gamma $. 
%For completeness, the curve representing the approach of Ref.~\cite{Molina:2019wjj} is also shown. In this last case $\chi^2 $ has been calculated only with the ratios $R_1, R_3, R_4, R_7 $, and the value of $\beta = 0.32 $ gives $\chi^2 \simeq 6 $. On the other hand, 
In the present estimation we include the following ratios with experimental data: $R_1, R_3, R_4, R_7 , R_8$ and $R_9$. We see that exclusion of $D$-wave mechanism yields $\chi^2 > 10^4 $ in the range of $\beta $ considered. This high value happens mainly due to the discrepancy between the respective theoretical results and experiments for the new ratios $ R_8$ and $R_9$ with only $S$-wave production mechanism. When the $D$-wave mechanism is taken into account, the $\chi^2 $ function has a significant diminution. Interestingly, for $\gamma = 10$ and $\beta = 0.32 $ we obtain $\chi^2 \simeq 11.8 $, but $\chi^2$ reaches its minimal value $ ( \approx 5.4 ) $ at $\beta \approx 0.05 $. In terms of the reduced $\chi^2 $ function,  $\chi^2/\nu $, with $\nu$ being the degree of freedom ($\nu =6-2=4$ in the present case), for $\beta = 0.32 $  we have $\chi^2/\nu  \approx 3.0 $,  and for  $\beta = 0.05 $,  $\chi^2/\nu  \approx 1.4 $. This is another way of verifying the relevance of the $D$-wave mechanism for tensor production.

Finally, we summarize in Table~\ref{tableratios} our results for the ratios taking $\gamma = 10$, with $\beta = 0.32$ and 0.05 for comparison. In the last column we also provide the respective experimental values when available.  In the end,  a fair accordance with experiments is obtained by providing a central place to the $D$-wave mechanism of tensor production. There seems to be a large discrepancy for $R_2$. The experimental ratio $R_2$ suffers from large uncertainties since the $ J/\psi \to \omega f_2^{\prime}(1525) $ decay is not observed and only a boundary is given. 
The fact that we obtain a reasonable result for $R_1$, which is the same ratio as $R_2 $ substituting the $ \omega $ by $ \phi $, should provide an incentive to measure the  $ J/\psi \to \omega f_2^{\prime}(1525) $ decay to see if the anomaly detected stands or not. 

The value of $\gamma \simeq 10$ needed to explain the data seems abnormally large. However, we already showed that the momentum of the odd vector, not forming the resonance, is very large, of the order of $1100 \, {\rm MeV} /c$. Assuming a range $R$ for the strong interaction of the order of 1 fm, the estimated angular momentum carried by this vector is $L = R . p = 5 - 6$. Hence, the $D$-wave component needed in our approach is not hindered by the kinematics of the reaction. Certainly, one could argue that the $J=0, J=2$ resonances that we obtain as dynamically generated could be more complicated than those provided by Refs.~\cite{Molina:2008jw,Geng:2008gx}, but the argument given above shows that it is not unreasonable to accept a large $D$-wave contribution in the reaction mechanisms discussed.

%%%%%%%%%%%%-Table1
\begin{table}[!]
%\footnotesize
\centering
\caption{Values obtained for the ratios defined in Eq.~(\ref{eq_ratios}) taking $\beta = 0.05, 0.32$,  $\gamma = 10$. In the last column we show the respective experimental values. }
\label{tableratios}
\setlength{\tabcolsep}{8pt}
\setstretch{1.2}
\begin{tabular}{lccc}
\hline 
  Ratio     & $\beta = 0.05 $ &  $\beta = 0.32 $ &  Experimental data \cite{Workman:2022ynf} \\
\hline
$R_1$	&	0.19	&	0.27 	& $0.4\pm 0.2$	 \\
$R_2$	&	2.07	&	1.43	& $>21 \pm 11$	 \\
$R_3$	&	13.78	&	10.07	& $13 \pm 3$	 \\
$R_4$	&	1.0	&	0.7 	& $1.1 \pm 0.2$	 \\
$R_5$	&	0.79	&	0.57	& $-$	 \\
$R_6$	&	0.013	&   0.033	& $-$	 \\
$R_7$	&	0.91 	&	1.24	& $1.3 \pm 0.4$	 \\
$R_8$	&	0.16	&	0.21	& $0.3	\pm 0.1$	 \\ 
$R_9$	&	2.47	&	1.74	& $3.1 \pm 0.6$		 \\ 
$R_{10}$	&	0.42	&	0.41	&  $-$		 \\
$R_{11}$	& 0.41		&	0.29	&  $-$		 \\
%\hline
\hline
\end{tabular}
\end{table}
%%%%%%%%%%%%-Table1

%%%%%%%%%%%%%%%%%%%%%%%%%%%%%%%%%%
%%%%%%%%%%%%%%%%%%%%%%%%%%%%%%%%%%
\section{Concluding remarks}            \label{sec:5}       %%%========SEC3
%%%%%%%%%%%%%%%%%%%%%%%%%%%%%%%%%%
%%%%%%%%%%%%%%%%%%%%%%%%%%%%%%%%%%

We have revisited the decay of the $J/\psi $ into an $\omega, \phi, K^{*0}$ and one of the $f_0(1370), f_0(1710), f_2(1270), f'_2(1525), K_0^{\ast}(1430)$ and $K_2^{\ast}(1430)$ resonances. We have taken advantage of previous works that considered this process as a $J/\psi $ decay into three vector mesons, and the interaction of two of them forming a scalar or tensor resonance. The purpose of the present work has been to extend these previous studies,  by using the same formalism concerning the flavor combinations, but investigating for the first time the relation between the scalar meson and tensor productions. To this end, we have performed an analysis of the spin structure of the four vectors present in the production vertex, and included the $D$-wave mechanism in the tensor production. This has allowed us to evaluate, beyond the ratios studied previously involving scalar states and tensor states independently, new ratios relating the scalar and tensor meson productions.

Our results suggest that a good agreement with experimental data is reached by adding a large contribution of the $D$-wave mechanism for the tensor production. At the best of our knowledge, unfortunately there are no yet experimental works reporting the angular distributions that might allow us to test this finding. In this sense, we hope that the experimentalist community also  revisits these reactions in order to get more and better statistical data samples. A partial wave analysis of the data, as done in most works, or the method of the moments of the distributions, which proves to be useful to separate the contribution of different angular momenta without the need to make a partial wave analysis~\cite{Du:2017zvv,Mathieu:2019fts,LHCb:2016lxy,LHCb:2014ioa,Bayar:2022wbx}, could be employed to clarify the issue.

%%%%%%%%%%%%%%%%%%%%%%%%%%%%%%%%%%
%%%%%%%%%%%%%%%%%%%%%%%%%%%%%%%%%%
\section*{Acknowledgments}
%%%%%%%%%%%%%%%%%%%%%%%%%%%%%%%%%%
%%%%%%%%%%%%%%%%%%%%%%%%%%%%%%%%%%

We would like to thank Jose Ramon Pelaez for valuable discussions concerning the large $N_c$ limit. L.M.A. has received funding from the Brazilian agencies Conselho Nacional
de Desenvolvimento Cient\'ifico e Tecnol\'ogico (CNPq) under contracts 309950/2020-1, 400215/2022-5, 200567/2022-5), Funda\c{c}\~ao de Amparo \`a Pesquisa do Estado da Bahia (FAPESB) under the contract INT0007/2016,  and CNPq/FAPERJ under the Project INCT-Física Nuclear e Aplicações (Contract No. 464898/2014-5). This work is partly supported by the National Natural Science Foundation of China under Grants Nos. 12175066,
11975009. This work is also partly supported by the Spanish Ministerio de
Economia y Competitividad (MINECO) and European FEDER funds under Contracts No. FIS2017-84038-C2-1-P
B, PID2020-112777GB-I00, and by Generalitat Valenciana under contract PROMETEO/2020/023. This project has
received funding from the European Union Horizon 2020 research and innovation programme under the program
H2020-INFRAIA-2018-1, grant agreement No. 824093 of the STRONG-2020 project.

%%%%%%%%%%%%%%%%%%%%%%%%%%%%%%%%%%
%%%%%%%%%%%%%%%%%%%%%%%%%%%%%%%%%%
%%%%%%%%%%%%%%%%%%%%%%%%%%%%%%%%%%
%=============== Refs ===============%  

\end{document}